\documentclass[conference]{IEEEtran}

\usepackage{cite}
\usepackage{amsmath,amssymb,amsfonts}
\usepackage{algorithmic}
\usepackage{graphicx}
\usepackage{textcomp}
\usepackage{xcolor}
\usepackage{booktabs} 
\usepackage{hyperref}

\def\BibTeX{{\rm B\kern-.05em{\sc i\kern-.025em b}\kern-.08em
    T\kern-.1667em\lower.7ex\hbox{E}\kern-.125emX}}

\makeatletter
\def\ps@IEEEtitlepagestyle{%
  \def\@oddfoot{\hfil\parbox{\textwidth}{\centering\small
  © This is the author’s version of the work published in the \textit{Proceedings of the IEEE/AIAA Digital Avionics Systems Conference (DASC) 2025}. DOI: 10.1109/DASC66011.2025.11257365.}\hfil}%
  
  \def\@evenfoot{}%
}
\makeatother

\begin{document}

\title{Quantum Enhanced Anomaly Detection for ADS-B Data using Hybrid Deep Learning}

\author{
    \IEEEauthorblockN{Rani Naaman\IEEEauthorrefmark{1}, 
                      Felipe Gohring de Magalhaes\IEEEauthorrefmark{1}, 
                      Jean-Yves Ouattara\IEEEauthorrefmark{1},
                      Gabriela Nicolescu\IEEEauthorrefmark{1}}
                      
    \IEEEauthorblockA{
        \IEEEauthorrefmark{1}Department of computer and software engineering, Polytechnique Montreal, Montreal (QC), Canada}
        
    \IEEEauthorblockA{Email: rani.naaman@polymtl.ca}
}

\maketitle

\begin{abstract}
The emerging field of Quantum Machine Learning (QML) has shown promising advantages in accelerating processing speed and effectively handling the high dimensionality associated with complex datasets. Quantum Computing (QC) enables more efficient data manipulation through the quantum properties of superposition and entanglement. In this paper, we present a novel approach combining quantum and classical machine learning techniques to explore the impact of quantum properties for anomaly detection in Automatic Dependent Surveillance-Broadcast (ADS-B) data. We compare the performance of a Hybrid-Fully Connected Quantum Neural Network (H-FQNN) with different loss functions and use a publicly available ADS-B dataset to evaluate the performance. The results demonstrate competitive performance in detecting anomalies, with accuracies ranging from 90.17\% to 94.05\%, comparable to the performance of a traditional Fully Connected Neural Network (FNN) model, which achieved accuracies between 91.50\% and 93.37\%.


\end{abstract}


\IEEEpubidadjcol



\section{Introduction}

Avionic systems have evolved significantly over the decades, transitioning from early federated architectures to today's advanced Integrated Modular Avionics (IMA) frameworks. In the original federated model, each aircraft function—such as navigation, communication, or flight control—was managed by independent hardware and software components, leading to increased weight, complexity, and maintenance challenges. Modern IMA architectures have revolutionized this landscape by centralizing processing power through shared computing resources and modular software platforms, greatly enhancing flexibility, scalability, and fault tolerance \cite{watkins2007ima}. This shift not only improves efficiency but also supports advanced communication protocols like Automatic Dependent Surveillance–Broadcast (ADS-B).\\


In recent years, the adoption of ADS-B systems has led the aviation industry to an increase in the volume of aircraft surveillance data. The ADS-B system provides real-time data by transmitting the aircraft' position, velocity and other flight characteristics to nearby aircraft and ground stations through a continuous, unencrypted data stream \cite{varga2015adsb}. By leveraging satellite navigation, the ADS-B acquires precise positional information of an aircraft and is designed to improve aviation safety by enhancing situational awareness for pilots and air traffic controllers, aiding in collision avoidance and preventing runway incursions \cite{manesh2017adsb}. Furthermore, it helps optimize routing, improving the overall efficiency of air traffic management. However, since aviation systems receive broadcast ADS-B data without validating the source or integrity, the system is vulnerable to cyber threats such as spoofing, ghost aircraft injection, trajectory manipulation, and other attacks through the transmission of false data into air traffic control systems without validation \cite{manesh2017adsb},\cite{kim2017adsb}.\\

Traditionally, a range of classical Machine Learning (ML) techniques have been employed in ADS-B security to detect anomalies and improve data validation. Algorithms such as Decision Trees, Support Vector Machines (SVM) and Neural Networks (NN) have demonstrated success in identifying and mitigating cyber threats in ADS-B data stream \cite{Cevi2024Anomaly},\cite{Pirolley2024GhostAircraft}. However, these classical methods often involve high computational complexity and memory usage and may lose critical temporal information during feature transformation. These limitations highlight the need for more adaptive and scalable approaches to secure ADS-B communication, especially as air traffic surveillance data grows in complexity and volume.\\

In parallel, as quantum hardware continues to improve, the growing field of QML has shown promising potential in accelerating the processing speed and effectively handling the high dimensionality associated with complex datasets through the quantum properties of superposition and entanglement \cite{Torabian2023Compositional},\cite{Hdaib2024}. Hybrid quantum-classical models, which combine quantum circuits with classical neural networks, offer the potential to enhance learning capabilities while remaining compatible with near-term noisy intermediate-scale quantum (NISQ) devices.\\

In this work, we explore the feasibility and application of a hybrid quantum-classical neural network, a H-FQNN, to perform anomaly detection on trajectory-based spoofing ADS-B data. We construct two quantum circuit-based hybrid models using one-hot encoded labels and binary cross-entropy loss and compare both to their classical counterpart, an FNN. The models are trained and evaluated on a publicly available ADS-B anomaly dataset that includes gradual manipulations of latitude and longitude, as well as merge attacks simulating new and complex flight patterns \cite{Cevi2024Anomaly}. Detecting such trajectory anomalies is critical in operational aviation environments, as they may lead to false traffic alerts and in-flight disruptions, potentially resulting in serious safety consequences \cite{Realities_adsb_martin}.\\

Notably, the experimental results provide empirical evidence of the feasibility and potential of QML in real-world aviation security use cases. With the inherent parallelism and potential speedup offered by QC, hybrid models display promising results for aviation safety, where every second counts in preventing threats. This research contributes to the ongoing exploration in safeguarding modern air traffic infrastructures and, to the best of our knowledge, is the first to apply QC to ADS-B data.\\



This paper is organized as follows: Section \ref{sec:basic} presents a review of the basic concepts on quantum computing, quantum machine learning and quantum circuits needed for the understanding of this paper's contribution. Section \ref{sec:sota} presents the related works and how the solution presented in this paper advances the limitations of the state-of-the-art. Section \ref{sec:meto} details the design of the models of H-FQNN and H-RNN, the preprocessing of the chosen dataset and the evaluation. Section \ref{sec:res} shows the results and a comparison to classical models followed by Section \ref{sec:dis}, where a discussion of results and the impacts of the model is presented. Further, it presents the limitations of current models and technology. Finally, Section \ref{sec:con} draws a conclusion to the paper and traces possible future work.
\section{Basic Concepts}
\label{sec:basic}
This section presents the basic concepts related to Quantum Computing and Quantum Machine Learning. The concepts serve as a base to the contributions presented in this paper, which rely on such new models.

\subsection{Quantum Computing (QC)}
Quantum computing can be understood as a computational paradigm that leverages the principles of quantum mechanics and is built upon the fundamental unit of quantum information, the qubit \cite{nielsen2010quantum}. Often represented as the spin of an electron or the polarization of a photon \cite{bernhardt2019quantum} \cite{hennig2024introducing}, the qubit is the unit of computation with the basis states of:
\begin{equation}
        \left| 0 \right\rangle = \begin{bmatrix} 1 \\ 0 \end{bmatrix}, \quad
        \left| 1 \right\rangle = \begin{bmatrix} 0 \\ 1 \end{bmatrix}
    \end{equation}

A multi-qubit quantum system can be expressed as the tensor product of individual qubit states. For example, the two-qubit state $|00\rangle$ is given by:
\begin{equation}
|00\rangle = |0\rangle \otimes |0\rangle = 
\begin{bmatrix} 1 \\ 0 \end{bmatrix} \otimes 
\begin{bmatrix} 1 \\ 0 \end{bmatrix} = 
\begin{bmatrix} 1 \\ 0 \\ 0 \\ 0 \end{bmatrix}
\end{equation}


The fundamental utility of the quantum systems lies in the ability to represent information using the quantum mechanical property of entanglement and superposition. Superposition is the property that enables a quantum system to exist in multiple states simultaneously until measured. This can be represented as the linear combinations of states, written $|\psi\rangle = \alpha |0\rangle + \beta |1\rangle$ where \( \alpha \) and \( \beta \) are complex coefficients for which the square of their magnitudes, \( |\alpha|^2 \) and \( |\beta|^2 \), give the probabilities of measuring the system in the \( |0\rangle \) and \( |1\rangle \) states. Entanglement is another elementary property defining a group of qubits which are interconnected and cannot be described independently. In a system of 2 entangled qubits, it can be interpreted as the state of one qubit directly influencing the state of the other, with the famous example of the Bell state \cite{bell1964einstein} \cite{qipc2021}. A simple representation of the equation is: 
\begin{equation}
|\psi\rangle = \frac{1}{\sqrt{2}} \left( |00\rangle + |11\rangle \right) .
\end{equation}

\subsection{Quantum Machine Learning (QML)}
\begin{figure}[t]
    \centering 
    \includegraphics[width=0.35\textwidth]{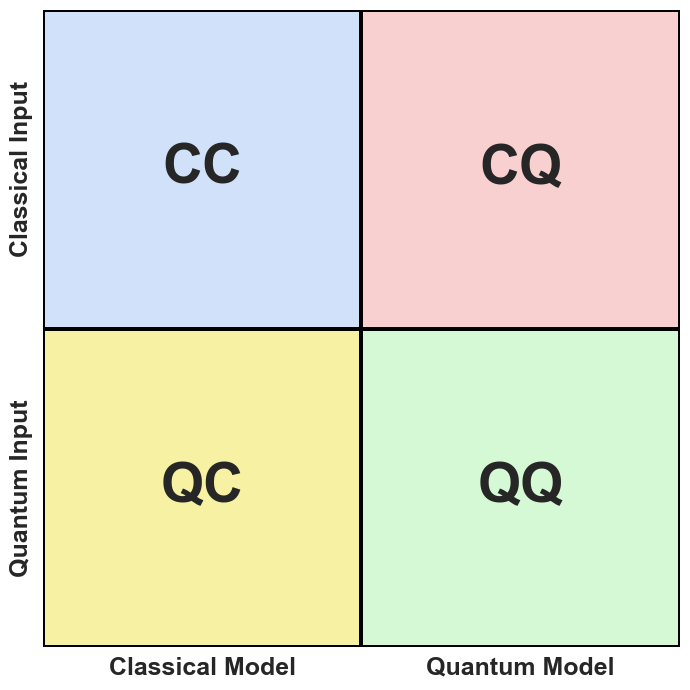} 
    \caption{Four different approaches to combine the disciplines of quantum computing and machine learning. (The figure is adapted from Schuld and Petruccione \cite{Schuld2021})} 
    \label{fig:4QC} 
\end{figure}

The convergence of Machine Learning and Quantum Computing has the potential to improve the communication technologies of the space and aeronautical sectors\cite{Brin2024}. Quantum Machine Learning brings together the current state of the art in Machine Learning and harnesses the quantum mechanical properties of entanglement and superposition to attempt to create better performing models. There are four different approaches to combine QC and ML \cite{Schuld2021} \cite{joshi2021evaluating} as illustrated in Figure~\ref{fig:4QC}. \texttt{Classical input-Classical model (CC)} refers to standard classical machine learning, where both data and models are classical. \texttt{Classical input-Quantum model (CQ)} involves processing classical data using quantum models, such as parameterized quantum circuits. \texttt{Quantum input-Classical model (QC)} describes the use of classical machine learning techniques to analyze data generated by quantum systems. Finally, \texttt{Quantum input-Quantum model (QQ)} leverages both quantum data and quantum models, enabling fully quantum learnable frameworks. These four paradigms define the various ways in which quantum and classical computing can synergize to advance machine learning capabilities. In this article, we refer to a \texttt{hybrid model} as a \texttt{CQ} framework that implements a quantum algorithm using classical input data.


\subsection{Variational Quantum Classifier (VQC)}
A Variational Quantum Classifier can be considered as the quantum equivalent of a layer in a hybrid NN model. VQCs are particularly well suited for NISQ computers as the process of optimizing the cost function of a machine learning algorithm also helps mitigate the inherent decoherence and errors present in current quantum hardware \cite{Havlicek2019}. The quantum circuit of a VQC begins with an initial state of qubits and uses a parameterized quantum circuit, known as a feature map, to encode features as quantum information \cite{Coles2021} \cite{Kruse2023Variational}. In this article, we used the angle embedding method, which encodes classical information of n features into the rotation angle of n qubits, effectively manipulating information from the vector space $\mathbb{C}^{N}$, where $N = 2^n$ is the dimension of the Hilbert space corresponding to n qubits \cite{Lloyd2020}. Specifically, we used \(R_X\) gates, which apply a rotation of $\theta$ around the x-axis of a qubit, mapping the classical feature $x \in \mathbb{R}$ to the quantum state $|\psi(x)\rangle \in \mathcal{H}$, where $\mathcal{H}$ is a two-dimensional Hilbert space. The \(R_X\) gate is defined by:

\begin{equation}
    R_X(\theta) = \begin{bmatrix} 
\cos\left(\frac{\theta}{2}\right) & -i\sin\left(\frac{\theta}{2}\right) \\
-i\sin\left(\frac{\theta}{2}\right) & \cos\left(\frac{\theta}{2}\right)
\end{bmatrix}
\end{equation}

Then, instead of using a fixed sequence of gates, a variational ansatz is used  to parameterize the quantum layer, with the parameters being optimized during the training process. We used strongly entangling circuits from Pennylane \cite{pennylane_strongly_entangling_layers}, which apply rotations around the Z, Y, and Z axes, followed by entangling CNOT gates \cite{schuld2018circuit}. The following gates are defined by:

\begin{equation}
    R_Y(\theta) = 
\begin{bmatrix}
\cos(\theta/2) & -\sin(\theta/2) \\
\sin(\theta/2) & \cos(\theta/2)
\end{bmatrix}
\end{equation}

\begin{equation}
    R_Z(\theta) = 
\begin{bmatrix}
e^{-i\theta/2} & 0 \\
0 & e^{i\theta/2}
\end{bmatrix}
\end{equation}

\begin{equation}
    \text{CNOT} = 
\begin{bmatrix}
1 & 0 & 0 & 0 \\
0 & 1 & 0 & 0 \\
0 & 0 & 0 & 1 \\
0 & 0 & 1 & 0
\end{bmatrix}
\end{equation}

For a single-qubit operation parameterized via Z-Y-Z rotations, the learnable unitary matrix is given by  \cite{nielsen2010quantum},\cite{Schuld2021}: 

\begin{equation}
\scalebox{0.80}{
$U(\theta_1, \theta_2, \theta_3) = 
\begin{bmatrix}
e^{-i(\theta_1 + \theta_3)/2} \cos(\theta_2/2) & -e^{-i(\theta_1 - \theta_3)/2} \sin(\theta_2/2) \\
e^{i(\theta_1 - \theta_3)/2} \sin(\theta_2/2) & e^{i(\theta_1 + \theta_3)/2} \cos(\theta_2/2)
\end{bmatrix}$
}
\end{equation}
where \( \theta_1, \theta_3 \) are \( Z \)-axis rotations and \( \theta_2 \) is a \( Y \)-axis rotation.\\


The qubit states are then decoded using expectation values in the Z basis, also known as the Pauli-Z measurement. This produces outcomes between $+1$ (when the state collapses to $|0\rangle$) and $-1$ (for $|1\rangle$). The expectation value $\langle Z \rangle$ and the Pauli-$Z$ matrix are defined as:
\begin{equation}
    \langle Z \rangle = \langle \psi | Z | \psi \rangle = |\alpha|^2 - |\beta|^2,
\end{equation}
where $|\psi\rangle = \alpha|0\rangle + \beta|1\rangle$ is the qubit state, and
\begin{equation}
    Z = 
    \begin{bmatrix}
        1 & 0 \\
        0 & -1
    \end{bmatrix}.
\end{equation}



\begin{figure}[t]
    \centering 
    \includegraphics[width=1.05\columnwidth]{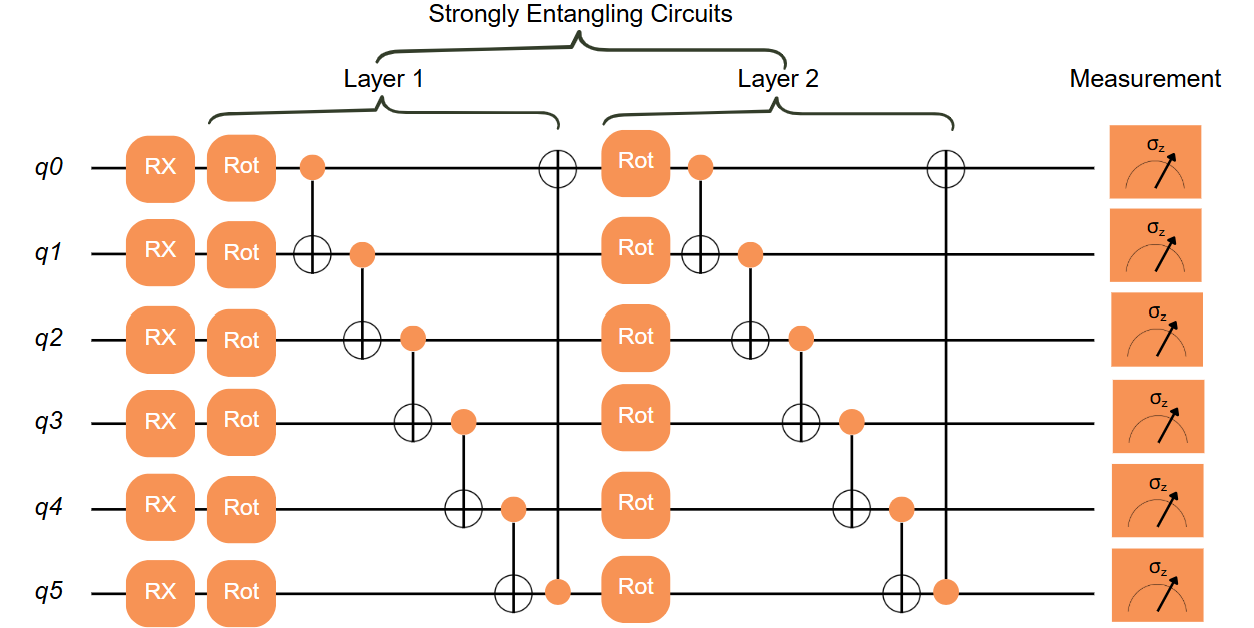} 
    \caption{An example of a VQC architecture with 6 qubits, strongly entangling layers and Pauli-Z basis measurement} 
    \label{fig:VQC} 
\end{figure}

By integrating a quantum feature map, a variational ansatz, and a measurement layer, we construct a VQC tailored for machine learning tasks. Figure \ref{fig:VQC} illustrates the architecture of a VQC with six qubits. The input data is encoded using \(R_X\) rotation gates, followed by two strongly entangling layers composed of learnable unitary rotation gates, defined as \(Rot\), and CNOT gates that entangle all six qubits. The circuit concludes with measurements in the Pauli-Z basis, represented by the standard measurement symbols (circle with an arrow).


\section{State-of-the-art}
\label{sec:sota}

Quantum Machine Learning is a relatively new field in the Internet of Things (IoT) domain, and its use in aeronautics is even more recent. Work from \cite{Templin2023Anomaly} highlighted the use of unsupervised deep generative models by implementing two variational autoencoders (VAE) models for anomaly detection in commercial flight operations: one utilizing a factorized Bernoulli prior and another employing a restricted Boltzmann machine (RBM). This study proposes an RBM model compatible with a quantum annealer or Quantum Circuit Born Machine (QCBM). The results demonstrate a mean precision of 0.591 for the RBM model, displaying competitive performance when compared to its Gaussian counterpart, but the performance needs further improvement to achieve reliability. While the technique used in this research differs from the current study, it demonstrates a progressive approach to applying QC to aeronautic data. In the paper by \cite{Gao2018IMA}, the authors present a deep learning method to evaluate the health state of the IMA. The QNN model is fed by extracted features from denoising autoencoders and classifies the data. Their method has been shown to be more effective and robust than the other four conventional algorithms, with an accuracy of 84.96\% in the testing data.\\

One notable work on the use of hybrid models for network security is proposed in \cite{Hdaib2024}, in which the authors develop three hybrid quantum-classical frameworks based on quantum autoencoders. These models aim to recognize network anomalies using quantum deep learning, demonstrating strong performance across standard datasets with NISQ-era IBM quantum simulators. According to their results, all three frameworks display promising detection performance. Notably, the integration of QAEs with Quantum K-Nearest Neighbor (QkNN) was shown to have the highest accuracy among the three. Moreover, the work from \cite{Cevi2024Anomaly} addresses the growing security concerns of ADS-B systems by proposing an anomaly detection system based on ML and DL techniques. In terms of results for their DL models, the multilayer perceptron achieved a competitive accuracy of 0.9817. In contrast to our work, which explores the potential of QML to handle anomaly detection in ADS-B data, the approach presented in \cite{Cevi2024Anomaly} mainly focuses on establishing a public ADS-B dataset and enhancing the performance of classical ML on that dataset.\\


While prior works have primarily focused either on classical machine learning techniques for avionics data or on applying QML to general network security challenges, our study bridges the gap between avionics, cybersecurity, and QML. By leveraging a hybrid quantum-classical model, we demonstrate the feasibility and effectiveness of quantum-enhanced models in detecting anomalies within  ADS-B flight data. This integration not only advances the application of QML in a domain with a high volume of data, but also sets a foundation for further exploration of quantum methods in real-time, safety-critical systems.


        

\section{Methodology}
\label{sec:meto}
This section outlines the design and implementation of our anomaly detection framework for ADS-B data using both classical and hybrid quantum-classical models. We begin by describing the architecture and design of both models. Following this, we detail the dataset preparation, including preprocessing and encoding steps. Finally, we outline the evaluation metrics used to benchmark the performance of the models. The goal is to establish similar steps between the classical and hybrid models to better compare the effects of QC, without introducing different biases.

\subsection{Model \#1: Hybrid-Fully Connected Quantum Neural Network (H-FQNN)}

\begin{figure}[t]
    \centering 
    \includegraphics[width=\columnwidth]{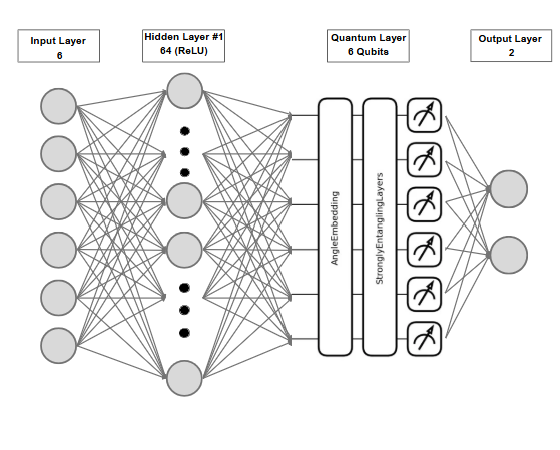} 
    \caption{Quantum Neural Network Architecture} 
    \label{fig:H-FQNN} 
\end{figure}

The H-FQNN model comprises four layers: three classical dense layers and one quantum layer. The first layer represents the input layer, which contains neurons equal to the number of features. The second layer is a hidden layer, also with neurons equal to the number of qubits, with a Rectified Linear Unit (ReLU) activation function. The third layer represents the quantum layer, also represented as a VQC. This quantum circuit contains qubits equal to the number of features, which are embedded with an R\_X gate, passed through strongly entangling circuits, and measured using Pauli Z operators. However, we also varied the number of qubits in this layer to explore the outcome on the performance. Then, the final layer is the output layer, which contains 2 neurons. As \texttt{BCEWithLogitsLoss} and \texttt{CrossEntropyLoss} include respectively a sigmoid and a softmax layer internally, the output layer does not have an activation function as it would result in redundant computation. Figure \ref{fig:H-FQNN} visually depicts the architecture of the H-FQNN. It is read from left to right, beginning with the classical input features, which are first processed by the classical layers and then passed to the quantum layer. The output of the quantum layer is then fed into the final dense layer for prediction, and the model is trained using the Adam optimizer \cite{kingma2017adammethodstochasticoptimization}.



\subsection{Model \#2: Fully Connected Neural Network (FNN)}
\begin{figure}[t]
    \centering 
    \includegraphics[width=0.42\textwidth]{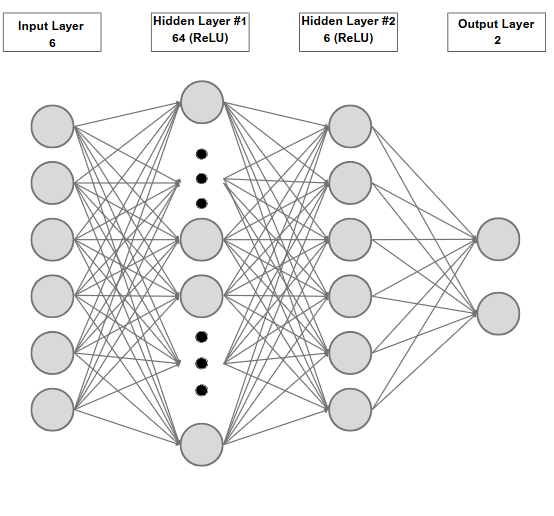} 
    \caption{Neural Network Architecture} 
    \label{fig:FNN} 
\end{figure}

The FNN model was developed to serve as a comparison against the hybrid model. This model shares a similar architecture with the hybrid model, consisting of four classical layers, and is trained using the Adam optimizer. However, the key difference lies in the third layer, which contains n neurons equal to the number of qubits with a ReLU activation function. Figure \ref{fig:FNN} illustrates the architecture of the FNN model, which is read in the same way as the H-FQNN model, but with a classical layer with six neurons replacing the quantum layer with six qubits.


\subsection{DATASET}







In aeronautic research, the availability and diversity of commercial aircraft data are challenges that can slow down novel advancements \cite{Li2019Reviewing}. To mitigate this difficulty and to focus on the impact of quantum properties, this research explored the Dataset 3 obtained from \cite{Cevi2024Anomaly}, which provided high-dimensional aircraft surveillance data for training and evaluation purposes. The dataset originates from the OpenSky Network, an open-access platform that provides ADS-B flight records \cite{Schaefer2014}. The dataset comprises essential flight parameters related to aircraft positioning and flight dynamics, including \texttt{latitude}, \texttt{longitude}, \texttt{barometric altitude}, \texttt{geometric altitude}, \texttt{velocity}, and \texttt{heading}. It also includes aircraft identifiers in the \texttt{ICAO24} format and a UNIX-formatted timestamps \texttt{time}. Dataset 3 was chosen among the available options because it contains basic trajectory manipulation attacks, including gradual drift in latitude/longitude as well as merge attacks that simulate complex flight patterns. Although these scenarios do not represent the full range of known ADS-B operational threats, the dataset provides an interpretable foundation for evaluating and isolating the contribution of QML methods in the study.


\subsection{Feature Selection}
Initially, the dataset relied on the eight flight parameters related to aircraft positioning and flight dynamics presented in Section IV.A. Prior to using the data, we performed feature selection to remove the parameters that negatively impacted the model's performance and processing efficiency. As the training of our H-FQNN did not use sequenced batches, we excluded the aircraft identification feature with the \texttt{ICAO24} format from the final model inputs to focus on learning general anomaly patterns. \\

\begin{figure}[t]
    \centering 
    \includegraphics[width=0.5\textwidth]{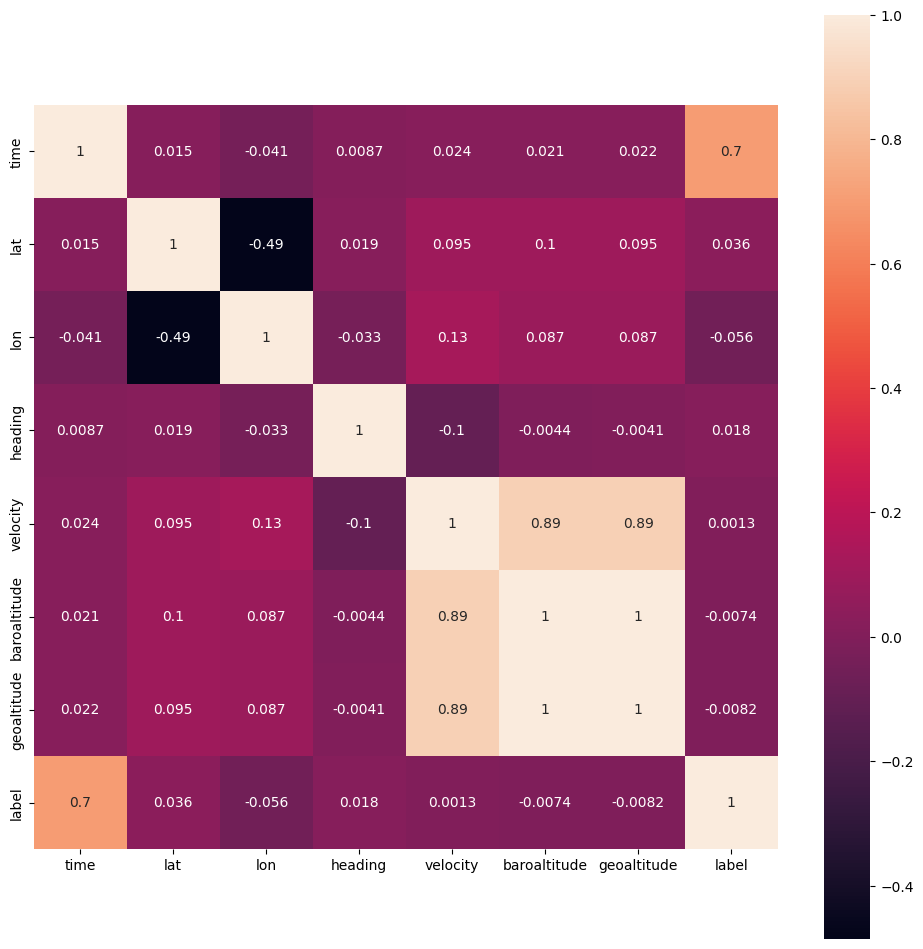} 
    \caption{Correlation matrix — Feature to label.} 
    \label{fig:Matrix} 
\end{figure}

Following the feature selection process, we examined pairwise relationships between flight parameters to identify and eliminate highly correlated features while preserving those most relevant for anomaly detection. The analysis process can be visualized through a complete correlation matrix represented in Figure \ref{fig:Matrix}. In this matrix, lighter colors indicate stronger linear correlations between features. The diagonal elements of the matrix represent the self-correlation of each feature, which is always equal to 1. The other elements show the correlation between pairs of features. From this full feature representation, a threshold was established at 90\% for parameter retention, eliminating strongly correlated variables while maintaining data integrity. We did not lower the threshold below 90\% to ensure the retention of the \texttt{velocity} feature and at least one \texttt{altitude} feature, which are critical for flight dynamics and aircraft positioning \cite{Cevi2024Anomaly}.  Notably, the features \texttt{barometric altitude} and \texttt{geoaltitude} displayed perfect collinearity. \texttt{barometric altitude} was then removed, retaining only the latter as our altitude reference. This filtering process effectively reduced the dataset to six essential features.


\subsection{Data Preprocessing}
\label{sec:preprocessing}


To prepare the ADS-B dataset for training and simulation, we applied a structured preprocessing pipeline, inspired from the work done in \cite{Payares2021Quantum}\cite{Hdaib2024}, composed of sampling, normalization, dimensionality reduction, and encoding to ensure compatibility with both classical and QML models. First, to maintain class balance and control dataset size, we performed stratified sampling by selecting a fixed number of attack and normal flight samples based on different ratios. This ensured representative subsets for training, validating and testing while using imbalanced dataset. The sampled dataset is represented as:

\begin{equation}
\mathcal{D}_{\text{sampled}} = \mathcal{D}_{\text{attack}}^{(n)} \cup \mathcal{D}_{\text{normal}}^{(rn)}
\end{equation}

where $\mathcal{D}$ is the dataset, $n$ is the number of selected attack samples and $r$ is the normal-to-attack ratio.\\

Next, all selected features were standardized using z-score normalization to ensure equal feature contribution:

\begin{equation}
x'_{ij} = \frac{x_{ij} - \mu_j}{\sigma_j}
\end{equation}

where $\mu_j$ and $\sigma_j$ denote the mean and standard deviation of feature $j$, respectively. This was implemented using \texttt{StandardScaler}. Standardization was applied on all models, since the first layer of both the H-FQNN and the FNN is a classical linear layer.\\

Finally, the dataset was partitioned using a stratified 70/20/10 ratio for training, validation, and testing. Two loss functions were explored and compared for comparison on quantum models. For the first loss function, \texttt{BCEWithLogitsLoss}, class labels were converted to one-hot encoded vectors. This binary vector format of 0s and 1s is compatible with the loss function \texttt{BCEWithLogitsLoss} and enables gradient updates in both classical and hybrid quantum-classical models. In contrast, \texttt{CrossEntropyLoss} was used with integer class labels instead of one-hot vectors. Feature matrices were converted to floating point tensors, while labels were stored as long integer tensors, as required by \texttt{CrossEntropyLoss}. This configuration allowed training with class indices, removing the need for one-hot encoding with \texttt{CrossEntropyLoss}.


\subsection{Evaluation Matrix}

\begin{figure}[t]
    \centering 
    \includegraphics[width=0.7\columnwidth]{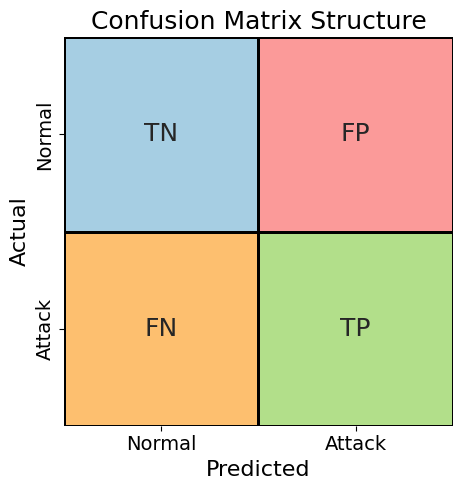} 
    \caption{Confusion matrix Structure} 
    \label{fig:Confusion Matrix} 
\end{figure}

The anomaly detection for ADS-B data was evaluated  using standard classification metrics, which compares the predicted classifications to the actual labels. The confusion matrix is a 2x2 table, shown in Figure \ref{fig:Confusion Matrix}, that summarizes the performance of a classification algorithm \cite{Ting2010}.

The corresponding definitions and evaluation metric are as follows:
\begin{itemize}
  \item \textbf{TP (True Positives)}: The number of samples correctly predicted as anomalies.
  \item \textbf{TN (True Negatives)}: The number of samples correctly predicted as normal.
  \item \textbf{FP (False Positives)}: The number of samples incorrectly predicted as anomalies when they are labeled as normal.
  \item \textbf{FN (False Negatives)}: The number of samples incorrectly predicted as normal when they are labeled as anomalies.
\end{itemize}

\begin{itemize}
    \item \textbf{Accuracy}:
    Proportion of correctly predicted anomalies and normal data:
    \begin{equation}
        \text{Accuracy} = \frac{TP + TN}{TP + TN + FP + FN}
    \end{equation}

    \item \textbf{Precision} :
    Proportion of true positives to the total number of predicted anomalies:
    \begin{equation}
        \text{Precision} = \frac{TP}{TP + FP}
    \end{equation}

    \item \textbf{Recall}: 
    Proportion of true positives to the total number of actual anomalies:
    \begin{equation}
        \text{Recall} = \frac{TP}{TP + FN}
    \end{equation}

    \item \textbf{F1 Score}:
    Combines precision and recall using their harmonic mean:
    \begin{equation}
        F_1 = 2 \times \frac{\text{Precision} \times \text{Recall}}{\text{Precision} + \text{Recall}}
    \end{equation}
\end{itemize}
\vspace{5 pt}
\section{Results}
\label{sec:res}

To validate our model, simulations were performed on a Lenovo Yoga Pro 9i equipped with an Intel Core i9-13905H processor, 32 GB of RAM and an NVIDIA GeForce RTX 4060 Laptop GPU. The Pennylane simulator, interfacing with Pytorch, was used to create quantum nodes. Furthermore, the classical models were executed on the GPU to accelerate the tests.  The tests were repeated multiple times to ensure robustness.\\

The results obtained in this study provide valuable insights into the performance of hybrid models in addressing ADS-B data, demonstrating the potential of QML for anomaly detection across varying hyperparameters. A summary of the performance metrics for both classical and quantum models is presented in Table \ref{tab:bce_Table} and Table \ref{tab:cross_Table}. Each row corresponds to a specific training scenario defined by the number of attack samples used during model training, and the results using the standard metrics.\\

\begin{table}[h]
\centering
\resizebox{\columnwidth}{!} {
\begin{tabular}{lcccccccc}
\toprule
\textbf{Model} & \textbf{Attack samples} & \textbf{Accuracy (\%)} & \textbf{Recall (\%)} & \textbf{Precision (\%)} & \textbf{F1 Score (\%)} \\
\midrule
FNN & 1000 & 92.67 & 91.00 & 92.63 & 92.61 \\
FNN & 5000 & 93.10 & 90.95 & 93.15 & 93.00 \\
FNN & 10000 & 93.05 & 90.59 & 93.19 & 92.92 \\
FNN & 20000 & 93.24 & 90.91 & 93.36 & 93.13 \\
FNN & 50000 & 93.30 & 91.04 & 93.40 & 93.19 \\
FNN & 75000 & 93.19 & 91.12 & 93.24 & 93.10 \\
\midrule
H-FQNN & 1000 & 92.00 & 91.50 & 92.11 & 92.04 \\
H-FQNN & 5000 & 93.37 & 91.45 & 93.39 & 93.29 \\
H-FQNN & 10000 & 94.05 & 92.44 & 94.05 & 93.99 \\
H-FQNN & 20000 & 93.20 & 91.39 & 93.20 & 93.13 \\
H-FQNN & 50000 & 93.14 & 91.12 & 93.17 & 93.05 \\
H-FQNN & 75000 & 93.40 & 91.65 & 93.39 & 93.33 \\
\bottomrule
\end{tabular}
}
\vspace{1pt}
\caption{Performance comparison of the FNN and H-FQNN  models with \texttt{BCEWithLogitsLoss} loss function and varying sample sizes. Training was performed using 150 epochs, a normal-to-attack ratio of 2 and a learning rate of 0.02.}
\label{tab:bce_Table}
\end{table}

\begin{table}[h]
\centering
\resizebox{\columnwidth}{!} {
\begin{tabular}{lcccccccc}
\toprule
\textbf{Model} & \textbf{Attack samples} & \textbf{Accuracy (\%)} & \textbf{Recall (\%)} & \textbf{Precision (\%)} & \textbf{F1 Score (\%)} \\
\midrule
FNN & 1000 & 91.50 & 91.38 & 91.77 & 91.57 \\
FNN & 5000 & 93.37 & 92.23 & 93.34 & 93.35 \\
FNN & 10000 & 92.80 & 91.93 & 92.80 & 92.80 \\
FNN & 20000 & 92.60 & 91.88 & 92.64 & 92.61 \\
FNN & 50000 & 92.13 & 91.61 & 92.23 & 92.16 \\
FNN & 75000 & 93.30 & 92.21 & 93.27 & 93.28 \\
\midrule
H-FQNN & 1000 & 90.17 & 91.25 & 91.23 & 90.34 \\
H-FQNN & 5000 & 92.43 & 91.30 & 92.41 & 92.42 \\
H-FQNN & 10000 & 92.95 & 91.89 & 92.93 & 92.94 \\
H-FQNN & 20000 & 93.25 & 92.68 & 93.30 & 93.27 \\
H-FQNN & 50000 & 92.81 & 92.43 & 92.92 & 92.84 \\
H-FQNN & 75000 & 92.74 & 92.48 & 92.89 & 92.79 \\
\bottomrule
\end{tabular}
}
\vspace{1pt}
\caption{Performance comparison of the FNN and H-FQNN models with \texttt{CrossEntropyLoss} loss function and varying sample sizes. Training was performed using 150 epochs, a normal-to-attack ratio of 2 and a learning rate of 0.02.}
\label{tab:cross_Table}
\end{table}


The results demonstrate that QML models can effectively process ADS-B data for anomaly detection. As shown in Table~\ref{tab:bce_Table} and Table~\ref{tab:cross_Table}, the H-FQNN models consistently achieved competitive performance across various dataset sizes, with accuracies ranging from 90.17\% to 94.05\%, and F1 scores reaching up to 93.99\%. The classical FNN models also performed strongly, achieving accuracies between 91.50\% and 93.37\%, and F1 scores up to 93.35\%. These findings confirm the feasibility of applying QML to aviation data, indicating that quantum circuits can be integrated into learning architectures for detecting anomalies in aircraft communication systems.\\


The comparison between FNN and H-FQNN reveals that QML can enhance detection accuracy in specific scenarios. For instance, using 5,000 attack samples and the \texttt{CrossEntropyLoss}, the H-FQNN outperformed the FNN with an accuracy of 92.43\% versus 93.37\%, and a nearly identical F1 score of 92.42\% compared to 93.35\%. With the \texttt{BCEWithLogitsLoss}, both models performed closely, with the H-FQNN achieving an F1 score of 93.29\%, compared to 93.00\% for the FNN. These results suggest that QML may offer marginal improvements in overall classification performance with the proposed architecture.\\

\begin{figure}[t]
\centering
\includegraphics[width=\columnwidth]{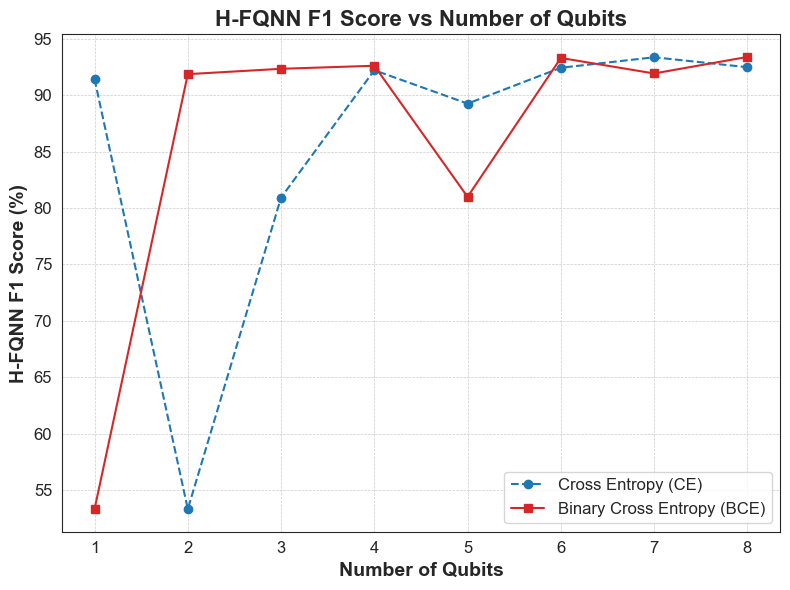}
\caption{F1 Score comparison between quantum models as a function of the number of qubits on the quantum layer. Training was performed using 150 epochs, a normal-to-attack ratio of 2, a learning rate of 0.02 and 5000 attack samples.}
\label{fig:accuracy_comparison}
\end{figure}

Figure~\ref{fig:accuracy_comparison} illustrates the variation of F1 scores with respect to the number of qubits used in the quantum circuit for both the \texttt{BCEWithLogitsLoss} and \texttt{CrossEntropyLoss} models. Both models exhibited peak performance around 6 qubits, with the \texttt{CrossEntropyLoss} model reaching an F1 score of 93.35\% and the \texttt{BCEWithLogitsLoss} model achieving 93.29\%. Although performance at seven and eight qubits appeared slightly higher in some instances, the marginal improvements did not justify the additional computational overhead. Therefore, six qubits were chosen as a practical balance between performance and efficiency for most simulations. These results emphasize the importance of quantum architecture tuning, as optimal performance was closely tied to the number of qubits used.\\



\begin{figure}[t]
\centering
\includegraphics[width=\columnwidth]{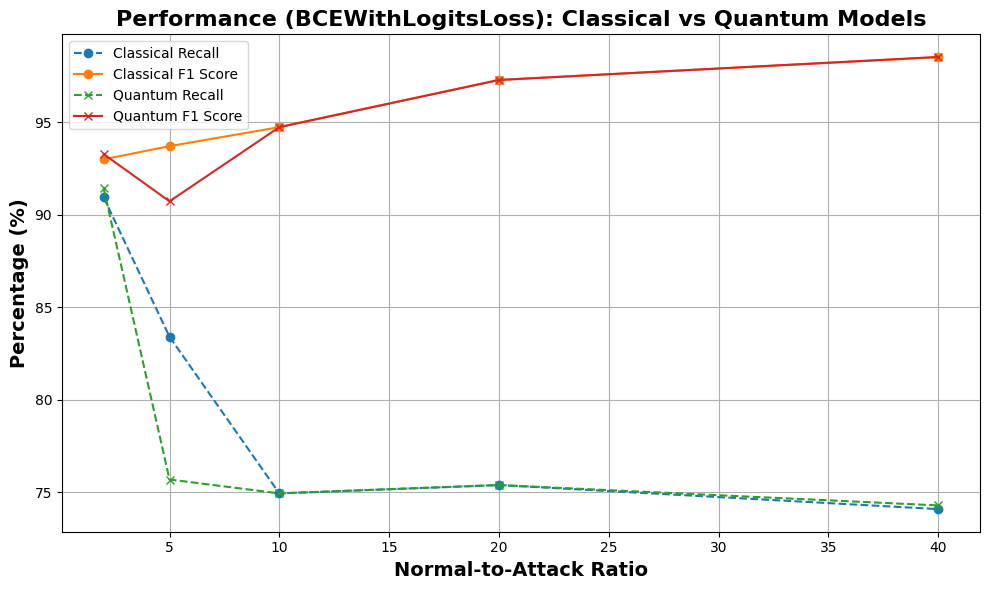}
\caption{Performance comparison of the FNN and H-FQNN models with varying normal-to-attack ratio using \texttt{BCEWithLogitsLoss}. Training was performed using 150 epochs, a learning rate of 0.02 and 5000 attack samples.}
\label{fig:bce_plot}
\end{figure}

\begin{figure}[t]
\centering
\includegraphics[width=\columnwidth]{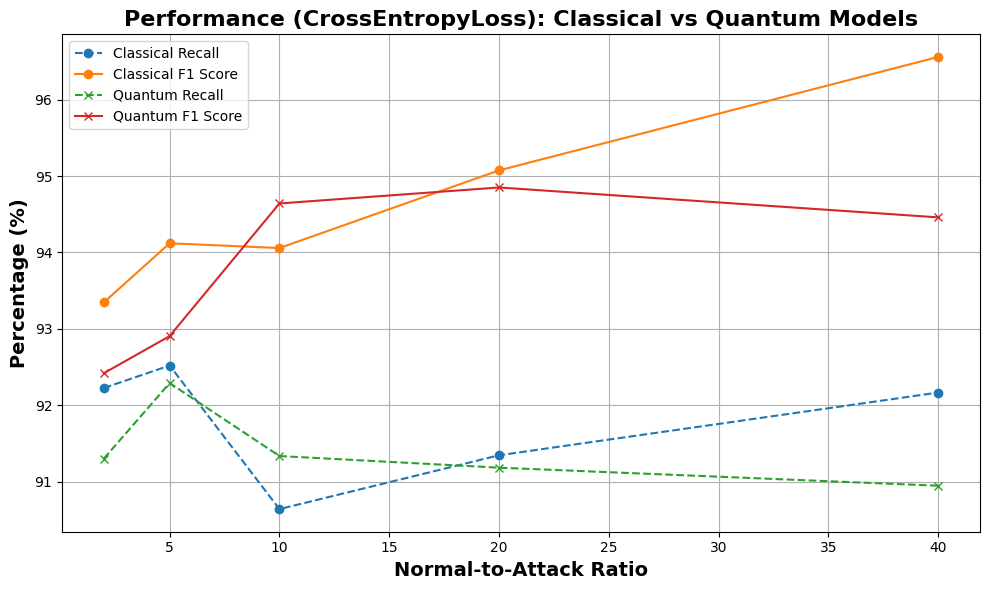}
\caption{Performance comparison of the FNN and H-FQNN models with varying normal-to-attack ratio using \texttt{CrossEntropyLoss}. Training was performed using 150 epochs, a learning rate of 0.02 and 5000 attack samples.}
\label{fig:ce_plot}
\end{figure}

The results presented in Figures~\ref{fig:bce_plot} and~\ref{fig:ce_plot} illustrate the performance of the classical and quantum models under varying normal-to-attack ratios using both \texttt{CrossEntropyLoss} and \texttt{BCEWithLogitsLoss}. At lower ratios, the classical model consistently demonstrates higher recall compared to the quantum model, with classical recall exceeding quantum recall by approximately 1–2\%. However, as the ratio increases, the performance gap between the models begins to narrow. Both the FNN and H-FQNN models exhibit improved F1 scores as the ratio grows, indicating better detection performance with more balanced datasets.\\

While the FNN maintains a slight edge in F1 score across all ratio levels, the H-FQNN remains competitive, especially at higher normal-to-attack ratios such as 20 and 40. At these levels, both models achieve strong recall and precision, but the FNN still shows slightly better balance between the two metrics. These findings suggest that classical models demonstrate stronger generalization in scenarios with imbalanced classes. Nevertheless, the quantum model yields promising results and continues to close the performance gap. Further optimization and tuning are necessary to fully leverage the potential of quantum enhanced models for aviation systems.

\section{Discussion}
\label{sec:dis}

We emphasize that the models in this study were developed with a focus on exploring the impact of quantum properties on anomaly detection on ADS-B data and not with the intention of creating the most effective models. The results from this study indicate that QML can effectively process ADS-B data for anomaly detection. The performance of the quantum models sometimes outperformed classical models, but it also performed slightly worse in other cases, indicating that hybrid models still require exploration and to establish an advantage over classical models. However, the models showed similar tendency and bias regarding the different techniques used. These important findings suggest that, when quantum devices will reach the Fault-Tolerant Quantum Computer (FTQC) era, the aeronautic domain could benefit from quantum properties to enhance its cybersecurity systems. Machine learning represents a prime candidate for quantum speedups as even polynomial improvements can yield substantial benefits due to the high computational demands of data processing \cite{Schuld2021}. As every second counts in aviation cybersecurity, the application of QML on the ADS-B is an excellent candidate.\\

 

It is also important to consider that all experiments were conducted on noise-free quantum simulators. The noise-free simulator from Pennylane does not account for real-world noise, decoherence, and gate errors inherent in quantum hardware. Moreover, running experiments on real quantum computers remains costly, which limits the feasibility of large-scale testing. While the impact of noise on current quantum computers is significant, the technology is continually advancing and new developments are improving hardware reliability and scalability. Relying on the IBM roadmap for quantum technologies, it is predicted that in the next 5 years, quantum computers will be more reliable in that sense \cite{ibm2024roadmap}. Accordingly, this work focused on the employment of QML, rather than on the impacts of imperfections on the quality of results. Beyond these general limitations, specific challenges arise from the nature of the project itself. The aerospace industry demands highly deterministic and reliable systems. Further exploration is needed to achieve reliable performance to align well with the strict safety requirements of aviation.\\
\section{Conclusion and Future Work}
\label{sec:con}

This study explored the application of QML for anomaly detection in ADS-B data. The findings demonstrated that H-FQNN models can effectively process ADS-B signals and achieve performance comparable to or exceeding that of classical models. Notably, the H-FQNN achieved a peak F1 score of 93.29\% using \texttt{BCEWithLogitsLoss} and 93.27\% using \texttt{CrossEntropyLoss}, both at six qubits equal to the number of features used. Although F1 scores remained high at seven and eight qubits, most simulations were limited to six qubits for practical considerations. These results suggest that quantum-enhanced architectures can effectively capture complex data relationships through quantum entanglement and superposition, indicating strong potential for applying QML to aviation anomaly detection tasks.\\


Moving forward, future research should focus on testing QML models on real quantum hardware, optimizing quantum circuit architectures for aerospace applications, and implementing more in-depth benchmarking against state-of-the-art classical anomaly detection models. While QML presents a promising direction for enhancing anomaly detection in aviation, further advancements in quantum computing hardware and algorithmic optimization are necessary to realize its full potential. By addressing these challenges, QML could contribute to more robust and efficient anomaly detection solutions for aircraft communication systems.



\bibliographystyle{ieeetr}

\bibliography{mybib}
\end{document}